\pdfoutput=1
\documentclass[a4paper]{article}
\usepackage[english]{babel}
\usepackage[utf8]{inputenc}
\usepackage{amsmath}
\usepackage[pdftex]{graphicx}
\usepackage[colorinlistoftodos]{todonotes}
\usepackage{authblk}
\usepackage[top=30truemm,bottom=30truemm,left=25truemm,right=25truemm]{geometry}
\usepackage{url}
\usepackage{cite}
\newcommand{\ds}{\displaystyle}
\usepackage{amssymb}
\usepackage{subcaption}

\title{Item Listing Optimization for E-commerce Websites \\ based on Diversity}
\author[1,2]{Naoki Nishimura\thanks{nishimura.n.ab@gmail.com}}
\author[2]{Kotaro Tanahashi}
\author[1,2]{Koji Suganuma}
\author[3,4]{\\Masamichi J. Miyama}
\author[3,4,5]{Masayuki Ohzeki}
\providecommand{\keywords}[1]{Keywords:}
\affil[1]{Internet Business Development Division, Recruit Lifestyle Co., Ltd., Tokyo, Japan}
\affil[2]{ICT Solution Department, Recruit Communications Co., Ltd., Tokyo, Japan}
\affil[3]{Graduate School of Information Sciences, Tohoku University, Sendai, Japan}
\affil[4]{$J_{ij}$ Inc., Tokyo, Japan}
\affil[5]{Institute of Innovative Research, Tokyo Institute of Technology, Yokohama, Japan}
\date{\vspace{-5mm}}

\begin{document}
\maketitle

\begin{abstract}
For e-commerce websites, deciding the manner in which items are listed on webpages is an important issue because it can dramatically affect item sales.
One of the simplest strategies of listing items to improve the overall sales is to do so in a descending order of sales or sales numbers.
However, in lists generated using this strategy, items with high similarity are often placed consecutively.
In other words, the generated item list might be biased toward a specific preference.
Therefore, this study employs penalties for items with high similarity being placed next to each other in the list and transforms the item listing problem to a quadratic assignment problem (QAP).
The QAP is well-known as an NP-hard problem that cannot be solved in polynomial time.
To solve the QAP, we employ quantum annealing (QA), which exploits the quantum tunneling effect to efficiently solve an optimization problem.
In addition, we propose a problem decomposition method based on the structure of the item listing problem because the quantum annealer we use (i.e., D-Wave 2000Q) has a limited number of quantum bits.
Our experimental results indicate that we can create an item list that considers both sales and diversity.
In addition, we observe that using the problem decomposition method based on a problem structure can lead to a better solution with the quantum annealer in comparison with the existing problem decomposition method.
\end{abstract}
\section{Introduction}
\begin{figure}[ht]
  \centering
  \includegraphics[width=8cm]{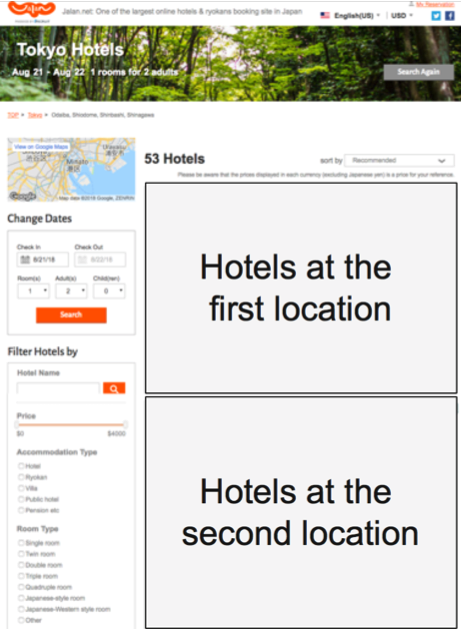}
\caption{An example of an item list on a hotel reservation website.}\label{fig:site_example}
\end{figure}

Several companies have recently started operating e-commerce websites to sell their items and services to the public considering the widespread use of the internet.
For these companies, deciding on the order in which items are listed on their website's pages is important because this ordering has the potential to dramatically affect the sales of their items or services.
Figure \ref{fig:site_example} shows a snapshot of a hotel reservation website. 
This is an example of the items being listed on an e-commerce page.
On this website, hotels at different locations are listed in the order of popularity from top to bottom.
These sorted items for display on webpages are collectively referred to as an item list.

To improve sales on e-commerce websites, placing items in the descending order of sales or sales numbers is a simple strategy for determining the list order of items \cite{long2014relevance}.
In addition, the sales of an item can be estimated when it is placed at different positions in the item list and the position of each product is determined to maximize the total sales estimate.
In particular, if $s_{ij}$ is the estimated sales of an item $i \in I$ when it is placed in a position $j \in J$, then the total sales for all items can be maximized by solving the following integer programming problem \cite{wang2016beyond}:

\begin{equation}
\begin{array}{|cll}
\ds \mathop{\mbox{maximize}}
& \ds \sum_{i \in I}\sum_{j \in J} s_{ij}x_{ij} \cr
\mbox{subject~to} 
& \ds \sum_{i \in I}x_{ij} = 1,~~j \in J, \cr
& \ds \sum_{j \in J}x_{ij} = 1,~~i \in I, \cr 
& \ds x_{ij} \in \{0,1\},~~i \in I,~~j \in J. \cr 
\end{array}
\label{form:ap}
\end{equation}
where $x_{ij}$ is a binary variable that indicates whether or not to assign item $i$ to position $j$. 
The abovementioned constraints ensure that only one item is allocated to each position, and only one position is allocated to each item.
In this study,

\begin{align}
S({\bf x})=\sum_{i \in I}\sum_{j \in J}s_{ij}x_{ij} \nonumber
\end{align}
is referred to as the sales term for ${\bf x}=(x_{11},x_{12},\cdots)$.
This problem can be interpreted as a network flow problem, and there exists an efficient technique to solve such a problem in polynomial time.
Furthermore, the solution obtained by solving this network flow problem with $x_{ij} \in [0,1]$ coincides with the solution of the abovementioned integer programming problem \cite{vazirani2013approximation}.

However, in the case of the list of items generated using such a strategy, the relationship between the different objects is ignored because the sales of each item $s_{ij}$ are considered independently.
For example, let us assume that customers visit an e-commerce website and browse the page of a particular item group.
If the relationships among different items are not considered while placing items in an item list, several items with high similarities can possibly be placed close to each other, thereby reducing the value of the item list for customers in terms of item diversity.
Considering this, several attempts have been made to include item diversity in item recommendation lists for users to ensure that they find the recommendation lists useful \cite{antikacioglu2017post,adomavicius2011maximizing}.
In these previous studies, the measures of diversity in the item recommendation lists for customers were improved by solving the maximum matching problem of the bipartite graph obtained after Top-$N$ recommendation.

The present study includes diversity in the item list targeting the entire user base using a proposed method of generating an item list by imposing penalties when items with a high similarity are placed at adjacent locations.
Thus, considering both sales and diversity, the item list generation problem can be formulated as a quadratic assignment problem (QAP) as detailed below.

We employ quantum annealing (QA) herein to solve the QAP \cite{Kadowaki1998}.
An optimization problem formulated with discrete variables can be efficiently solved using the Ising model or a quadratic unconstrained binary optimization problem (QUBO) because of the introduction of the quantum tunneling effect by QA.
Currently, the protocol of QA is artificially realized in an actual quantum device known as a quantum annealer \cite{Dwave2010a,Dwave2010b,Dwave2010c,Dwave2014}.
The quantum annealer has been tested for numerous applications, including portfolio optimization \cite{Rosenberg2016}, protein folding simulation  \cite{Perdomo2012}, molecular similarity problem \cite{Hernandez2017}, computational biology \cite{Richard2018}, job-shop scheduling \cite{Venturelli2015}, traffic optimization \cite{Neukart2017}, election forecasting \cite{Henderson2018}, machine learning \cite{Crawford2016,Neukart2018,Khoshaman2018}, and for automated guided vehicles in plants \cite{Ohzeki2019}.
In addition, several other studies have been conducted to efficiently solve various problems using the quantum annealer \cite{Arai2018nn,Takahashi2018,Ohzeki2018NOLTA,Ohzeki2018,Okada2019}.

In particular, our problem can be solved using the quantum annealer by formulating our QAP as a QUBO.
However, a QUBO for such a large number of items cannot be directly solved in one instance with the current state-of-the-art quantum annealer, namely D-Wave 2000Q, because the D-Wave 2000Q employs the chimera graph. 
The physical qubits available on D-Wave 2000Q are less than $2048$ because the qubits might have defects. 
In addition, the connection between the physical qubits is sparse and limited on the chimera graph.
Thus, several embedding techniques have been proposed; however, the number of logical qubits available to represent the optimization problems to be solved is drastically reduced \cite{boothby2016fast}.
To address this issue, a heuristic method has been proposed to solve a large-sized problem using a limited number of hardware bits.
D-Wave Systems, which is the manufacturer of D-Wave 2000Q, has developed an open-source software \texttt{qbsolv} \cite{booth2017partitioning} that solves a large-sized problem by dividing it into small subproblems.
However, the decomposed QAP might not necessarily lead to feasible solutions because \texttt{qbsolv} selects the subset of variables in the order of the energy impact of each variable for division of a problem.
Furthermore, in the literature \cite{Okada2019}, the division of an original problem into subproblems based on its structure is a promising method to efficiently solve large-sized problems using D-Wave 2000Q.
Considering this, we propose herein a method of obtaining better objective values for the QAP problem compared to those of the existing method \cite{booth2017partitioning} in the same calculation time by decomposing the problem based on the set of items and positions that they can be assigned to in an item list.
In addition, we assess the performance of the proposed method using the actual access log of a hotel reservation website.

The primary contributions of our study are summarized as follows:
\begin{itemize}
\item We propose a method of creating item lists on an e-commerce website as a QAP considering the sales and diversity of the items.
\item We convert the QAP to the QUBO to solve the abovementioned problem with D-wave 2000Q.
\item We propose a decomposition technique exploiting the structure of the item list.
\end{itemize}

\section{Models}\label{sec:models}
\subsection{Formulating the Item Listing Optimization Problem as a QAP}
We introduce the diversity term in our proposed model to add diversity in the item list.
We particularly calculate the similarity $f_{ii'}$ for pairs of items $i, i' \in I$ to introduce diversity in the item list.
The diversity of the item list (i.e., the diversity term) is defined as the negative value of the summation of the items’ similarity degree $f_{ii'}$ for overall adjacent items:

\begin{align}
D({\bf x})= -\sum_{i \in I}\sum_{i' \in I}\sum_{j \in J}\sum_{j' \in J}f_{ii'}d_{jj'}x_{ij}x_{i'j'} \nonumber \label{diversity_term} .
\end{align}
where $d_{jj'}$ is the adjacent flag of the position $j$ and $j'$; $d_{jj'}=1$ is for the adjacent positions; and $d_{jj'}=0$ is for the non-adjacent positions.
The value of function $D({\bf x})$ decreases because the high-similarity items are adjacent.
We solve a multi-objective optimization problem based on two values: the sales of individual products $S({\bf x})$ and diversity of the item list $D({\bf x})$.
Let $w$ be a parameter used to determine the penalty for listing items with high similarity. This problem is formulated as follows:

\begin{equation}
\begin{array}{|cll}
\ds \mathop{\mbox{maximize}}
& \ds \sum_{i \in I}\sum_{j \in J} s_{ij}x_{ij} - w \sum_{i \in I}\sum_{i' \in I}\sum_{j \in J}\sum_{j' \in J}f_{ii'}d_{jj'}x_{ij}x_{i'j'} \cr
\mbox{subject~to} 
& \ds \sum_{i \in I}x_{ij} = 1,~~j \in J, \cr
& \ds \sum_{j \in J}x_{ij} = 1,~~i \in I, \cr 
& \ds x_{ij} \in \{0,1\},~~i \in I,~~j \in J. \cr 
\end{array}
\label{form:qap}
\end{equation}
Two methods for calculating the similarity $f_{ii'}$ are introduced: explicit and implicit expressions.
In the explicit expression, semantic features, such as product category and average price, can be quantified using the distances between the feature vectors that can be calculated. The smaller the feature distance, the higher the similarity $f_{ii'}$ between the items.
In the implicit expression, the higher the co-browsing number of an item (i.e., the number of times that the items were viewed in the same session by the same user on the website), the higher the similarity $f_{ii'}$.

The advantage of the first approach is that the interpretation of the result is straightforward; nevertheless, it suffers from a disadvantage in that appropriate semantic features must be created and quantified.
In contrast, the advantage of the second approach is that it involves easy calculations and can consider various information reflecting customer behavior; however, its disadvantage is that semantic interpretation might be difficult.
Nevertheless, we employed herein the latter method of counting the co-browsing number to calculate similarity $f_{ii'}$.
Section \ref{sec:ex_setup} describes the detailed calculation for this.
As previously specified, the optimization problem (\ref{form:qap}) is a QAP.

The QAP is well-known as an NP-hard problem that cannot be solved in polynomial time \cite{anstreicher2003recent, abdel2018comprehensive}.

\subsection{Formulating the Item Listing Optimization Problem as a QUBO}
We utilize QA to solve our optimization problem.
The optimization problem must be expressed in the form of a QUBO to use QA as a solver. QUBO is given as follows \cite{Lucas2014}:

\begin{equation}
\begin{array}{|cll}
\ds \mathop{\mbox{minimize}}
& \ds {\bf x}^{\mathrm{T}} Q {\bf x} \cr 
\mbox{subject~to} 
& \ds {\bf x} \in \{0,1\}^N, \cr 
\end{array}
\label{form:qubo}
\end{equation}
where $Q \in \mathbb{R}^{N \times N}$.
Thus, our optimization problem can be transformed into a QUBO by employing a penalty function for violating constraints and adding this penalty function to the objective function:
\begin{equation}
\begin{array}{|cll}
\ds \mathop{\mbox{minimize}}
& \ds -\sum_{i \in I}\sum_{j \in J} s_{ij}x_{ij} + w \sum_{i \in I}\sum_{i' \in I}\sum_{j \in J}\sum_{j' \in J}f_{ii'}d_{jj'}x_{ij}x_{i'j'} \cr
& \ds +M \left( \sum_{i \in I}\left(\sum_{j \in J}x_{ij}-1\right)^2  + \sum_{j \in J} \left( \sum_{i \in I}x_{ij}-1 \right)^2 \right) \cr
\mbox{subject~to} 
& \ds x_{ij} \in \{0,1\},~~i \in I,~~j \in J. \cr 
\end{array}
\label{form:qap_qubo}
\end{equation}
where $M$ is a parameter used to prevent the violation of the constraint conditions. 
This is ensured by setting an appropriate value for $M$.
In theory, $M$ should take an extremely large value.
However, we cannot set $M$ to such a large value because of the limitations of the current version of the quantum annealer used (i.e., D-Wave 2000Q).
Thus, for simplicity, we present $M$ to the size of the largest element of the absolute value of $Q$ in (\ref{form:qubo}).

\subsection{Decomposition Methods for Item Listing Problems}
\texttt{qbsolv} is a software tool released by D-Wave Systems that enables solving a QUBO larger than one that can be processed using D-Wave 2000Q \cite{dwave2017solver}. \texttt{qbsolv} is essentially a decomposing solver that divides a large problem into smaller parts, which can then be solved by D-Wave 2000Q.
Thus, when a large QUBO is inputted, \texttt{qbsolv} divides the problem and sends each part of the problem independently to D-Wave 2000Q for calculation to obtain partial solutions.
This process is repeated by selecting different parts of the problem using the tabu search until solution improvement stops.
See \cite{booth2017partitioning} for the detailed algorithm of \texttt{qbsolv}.
Furthermore, \texttt{qbsolv} selects the subset of variables in the order of the energy impact of each variable for division of a problem.
However, in some cases, no feasible solution can be obtained when the target variables are extracted, regardless of the structure of the original problem.

\begin{figure}[ht]
\centering
  \begin{minipage}[b]{0.45\linewidth}
    \centering
    \includegraphics[width=8cm]{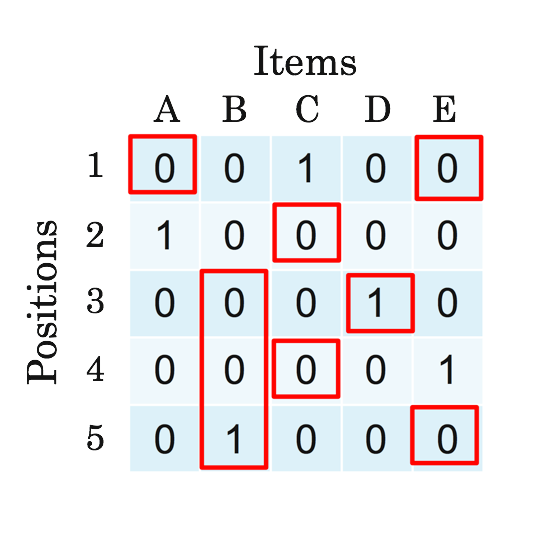}
  \end{minipage}
  \begin{minipage}[b]{0.45\linewidth}
    \centering
    \includegraphics[width=8cm]{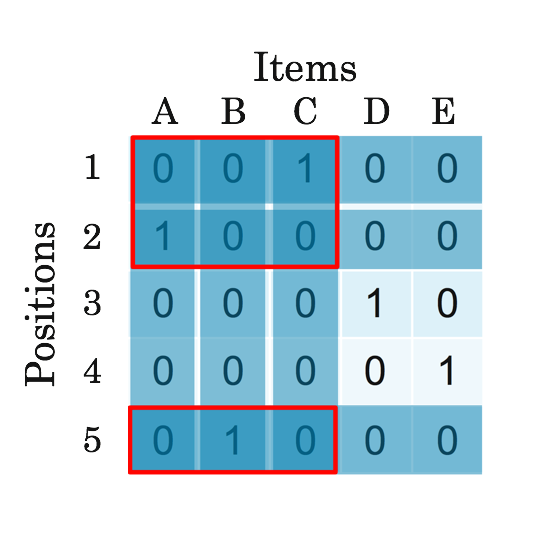}
  \end{minipage}
  \caption{Example of problem decomposition. The red frames represent the variables of the subproblem to be selected. The left figure is a selection of a subproblem without considering the problem structure. A subproblem has no feasible solution. The right figure is a selection of a subproblem based on the logical structure of the problem. A subproblem has feasible solutions.}\label{fig:decomp}
\end{figure}

Therefore, we focus herein on the structure of the assignment problem and propose a method to extract problems with feasible solutions.
Particularly in the case of an assignment problem, one condition involves each item being necessarily assigned to one position and another condition, in which each position is necessarily assigned to one item.
Therefore, while dividing the problem, we have to select variables with candidate combinations of items and positions that are already assigned.
Figure \ref{fig:decomp} shows an example of the decomposition.

The original problem can be decomposed as follows if the number of items in the original problem is $N_{{\rm org}}$ and the number of items solved by a partial problem is $N_{\rm sub}$:

\begin{enumerate}
\item Let $\mathcal{N}_s$ be the set of $N_{\rm sub}$ items extracted from $N_{\rm org}$ items.
\item Let $\mathcal{P}_s$ be the set of positions of items of $\mathcal{N}_s$.
\item Let $\mathcal{N}_s \times \mathcal{P}_s$ be the variables of the decomposed problem.
\end{enumerate}
This procedure involves ${}_{N_{\rm org}^2} \mathrm{C} {}_{N_{\rm sub}^2}$ candidates for variable combinations in the selection of the subproblems; however, the number of solution candidates can be reduced to ${}_{N_{\rm org}} \mathrm{C} {}_{N_{\rm sub}}$ exploiting the structure of the item list.

In practice, it is most important to determine the order in which items are listed in the upper positions of the item list because they are the items that are browsed most often.
Therefore, it is effective to solve the entire list as an integer programming problem as in Problem (\ref{form:ap}) first, then only resolve the particularly important upper positions of the list using the QAP (\ref{form:qap}).

\section{Results and Discussion}
\subsection{Experimental Setup}\label{sec:ex_setup}
For our experiments, we used the actual access log data of the online hotel reservation site Jalan \footnote{\url{ https://www.jalan.net/en/japan_hotels_ryokan/}}.
On this e-commerce website, a hotel list is created daily based on each area in Japan and the number of guests, including adults and children, that the hotels can accommodate in their rooms.
The access log includes the date and the time the customer accessed the item list screen, position of each item when the item list screen was accessed, and information on the hotel at which the customer made a reservation.
We estimated the sales $s_{ij}$ and similarity $f_{ii'}$ for the top 10 accessed areas on the hotel reservation website using the access log for the past six months.
Various methods can be considered for estimating $s_{ij}$.
The similarity $f_{ii'}$ was estimated using the log of the co-browsed items in the same session of the customer.
Furthermore, $s_{ij}$ and $f_{ii'}$ were normalized such that their average was 0, and the standard deviation was 1 for each item list.

We conducted two experiments in this study:
\begin{itemize}
\item evaluating the effect of the diversity term: comparison of solutions when the diversity control parameter is changed for the QAP (\ref{form:qap});
\item
performance evaluation of problem decomposition: comparison of the objective values when the structure of the item list is considered for the \texttt{qbsolv} problem decomposition.
\end{itemize}
As previously specified, we used D-Wave 2000Q (DW\_2000Q\_VFYC\_2) for our experiments.
Coupler strengths mapping logical to physical couplers with two physical couplers connecting each pair of logical qubits were set as 3.0.
Table \ref{tab:decomp_parameter} lists the values set for the parameters of D-Wave 2000Q and \texttt{qbsolv}.
\texttt{num\_read}, \texttt{annealing\_time}, and \texttt{repeats} represent the number of requests for problems, time per annealing, and number of times the main loop of the algorithm is repeated with no change in the optimal value before stopping, respectively. \cite{dwave2017solver,booth2017partitioning} 

\begin{table}[ht]
\centering
\caption{Parameters used for solving the problem in our numerical experiments.}\label{tab:decomp_parameter}
\begin{tabular}{cc} \hline
Parameter & Value \\ \hline
\texttt{num\_reads} \cite{dwave2017solver}  & 1000 \\
\texttt{annealing\_time} \cite{dwave2017solver}  & 20~[$\mu$s] \\
\texttt{auto\_scale} \cite{dwave2017solver}  & True \\
\texttt{postprocess} \cite{dwave2017solver}  & optimization \\
\texttt{num\_spin\_reversal\_transforms} \cite{dwave2017solver}  & 4 \\
\texttt{timeout} \cite{booth2017partitioning}  & 20~[s]  \\
\texttt{repeats} \cite{booth2017partitioning}  & 5 \\
\texttt{subproblemSize} \cite{booth2017partitioning}  & 64 \\ \hline
\end{tabular}
\end{table}

\subsection{Effect of the Diversity Term}
D-Wave 2000Q has less than $2048$ qubits because the qubits typically have defects. In addition, as previously specified, the connection between the physical qubits is sparse and limited on the chimera graph, in which D-Wave 2000Q has been based on. Thus, we can consider the problem with eight items per subproblem because complete graph embedding can be applied to arbitrary problem graphs with less than 64 logical variables.
Therefore, we first compare the sales term $S({\bf x})$ and the diversity term $D({\bf x})$ obtained by solving the QAP (\ref{form:qap}) by changing the diversity control parameter $w$ in the problem when the number of items is 6 and 8.
The obtained solution will be the same as that obtained via solving (\ref{form:ap}) if $w$ is $0$.

\begin{figure}[ht]
\centering
  \begin{minipage}[b]{0.45\linewidth}
    \centering
    \includegraphics[width=7.6cm]{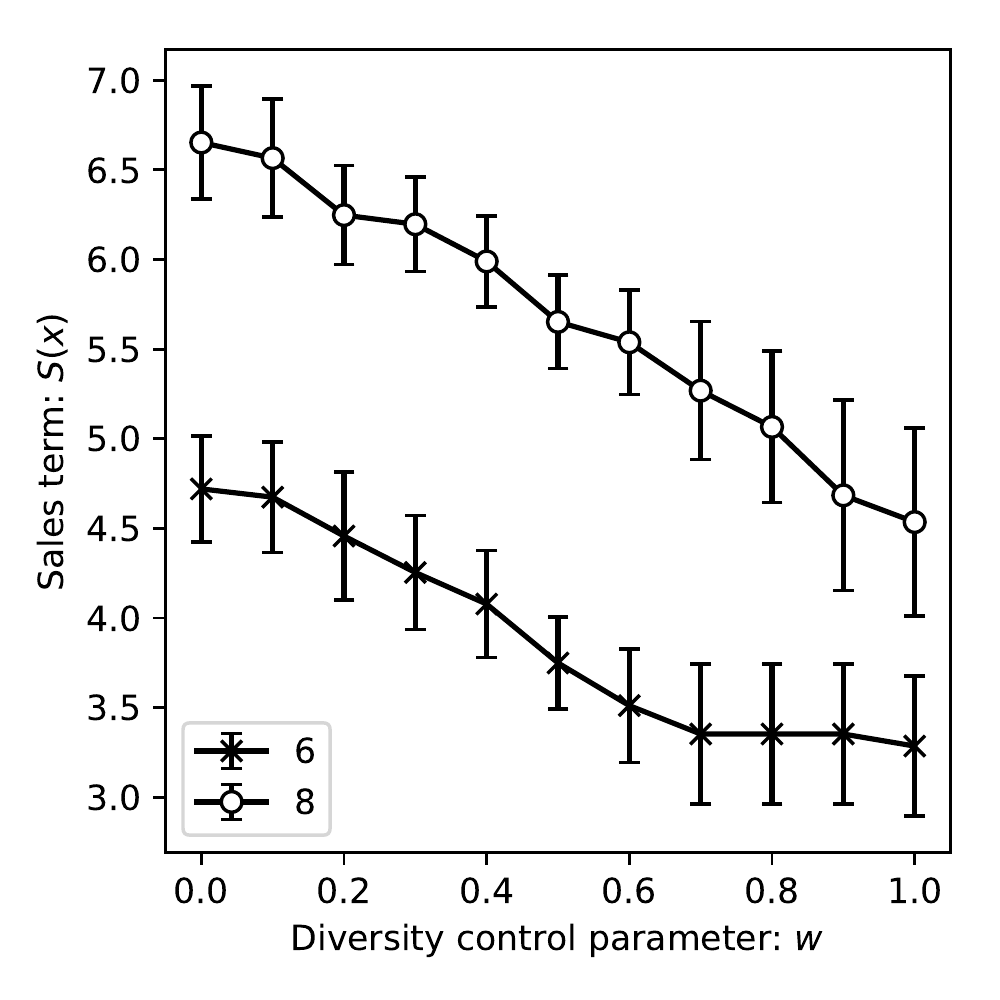}
    \subcaption{The average sales term value of 10 areas change when the number of items is changed.}\label{sales_average}
  \end{minipage} 
  \begin{minipage}[b]{0.45\linewidth}
    \centering
    \includegraphics[width=7.6cm]{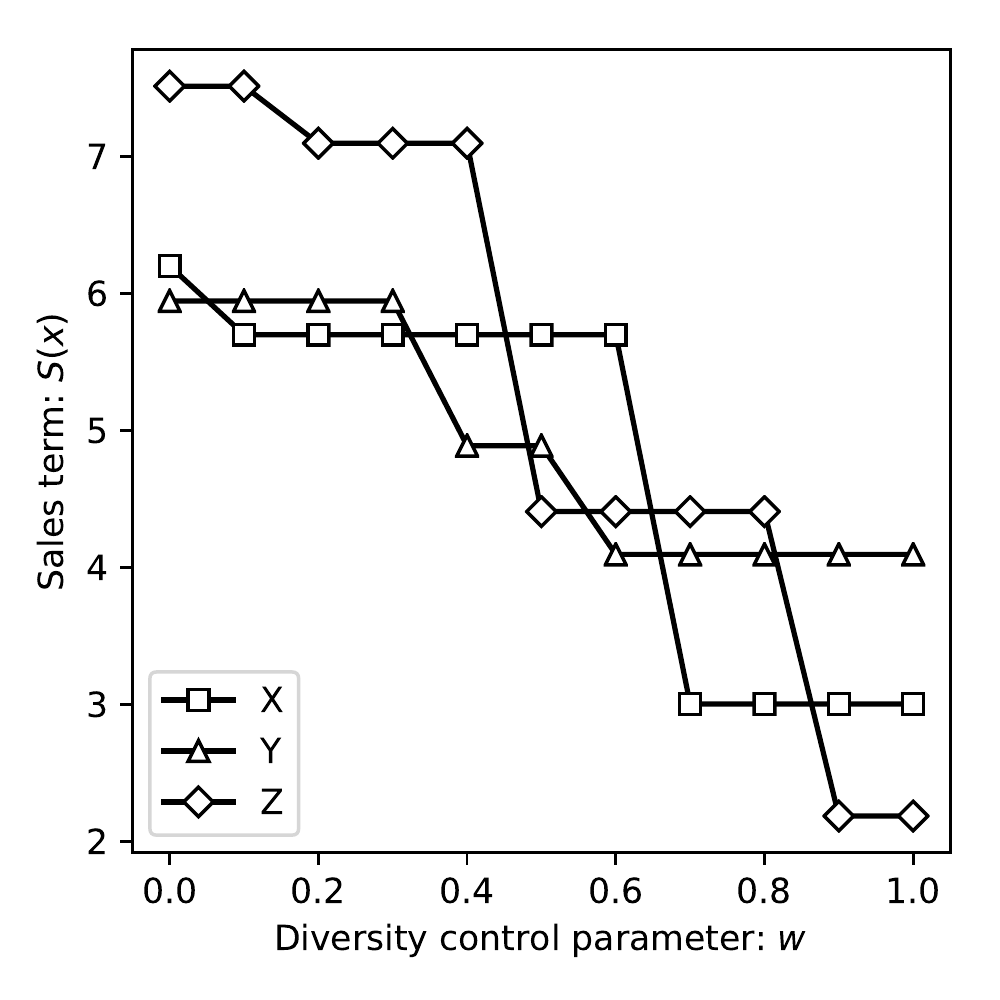}
    \subcaption{Changes in the typical sales term when the number of items is 8.}\label{sales_typical}
  \end{minipage}
  \caption{Changes in the sales term when the diversity control parameter is changed.}\label{fig:sales}
\end{figure}

\begin{figure}[ht!]
\centering
  \begin{minipage}[b]{0.45\linewidth}
    \centering
    \includegraphics[width=7.6cm]{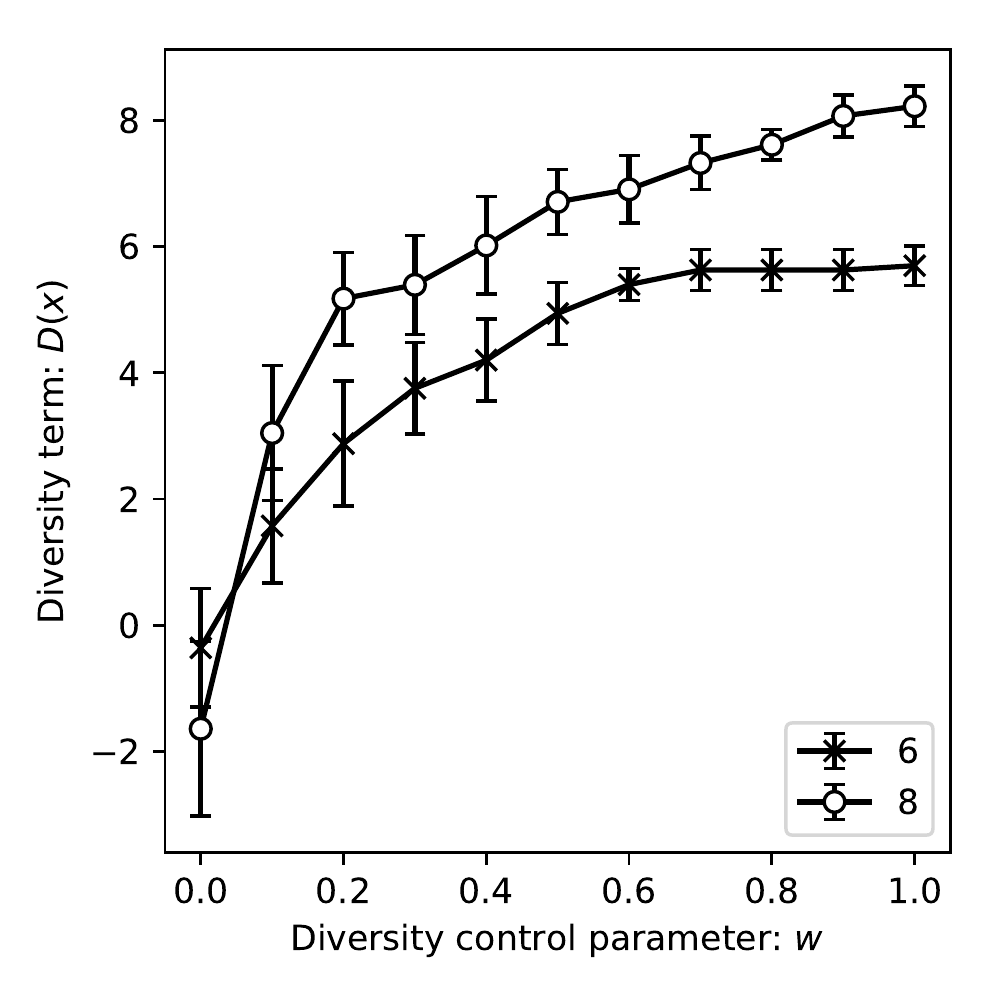}
    \subcaption{The average diversity term of 10 areas change when the number of items is changed.}\label{diversity_average}
  \end{minipage} 
  \begin{minipage}[b]{0.45\linewidth}
    \centering
    \includegraphics[width=7.6cm]{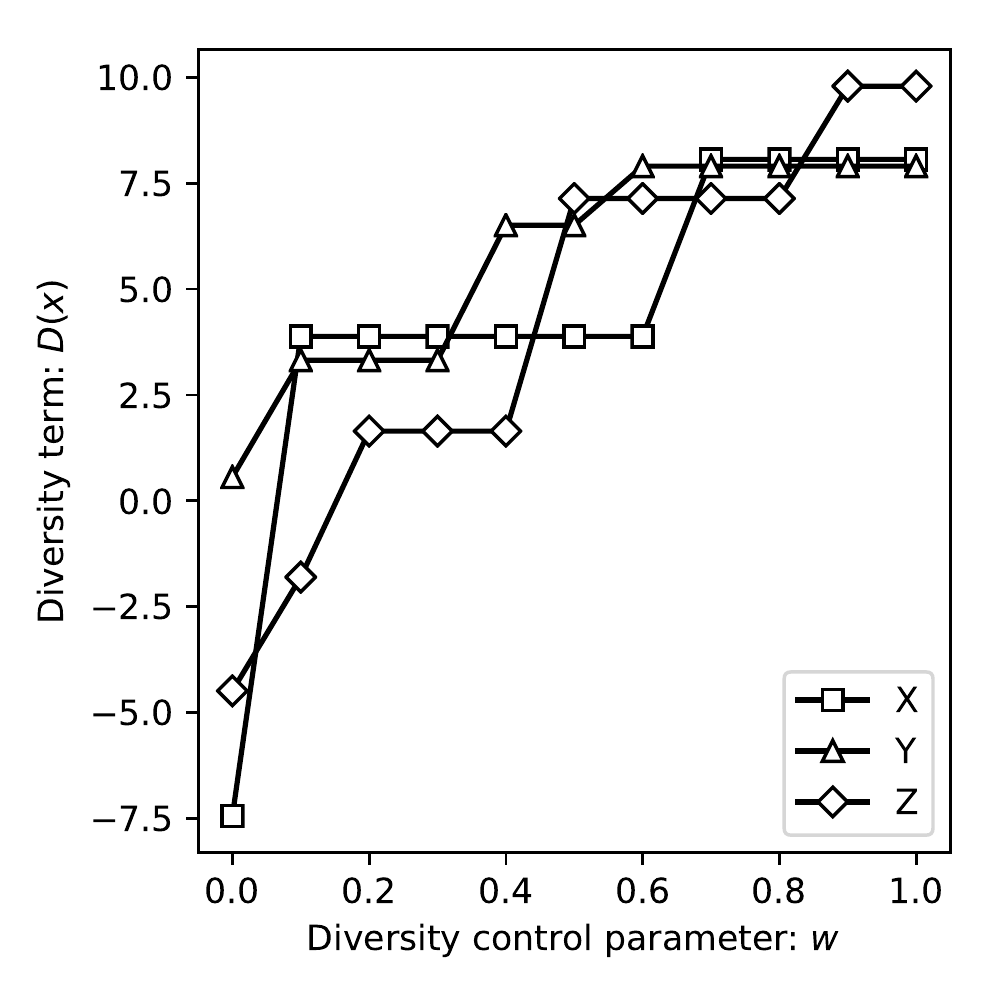}
    \subcaption{Changes in the typical diversity term when the number of items is 8.}\label{diversity_typical}
  \end{minipage}
  \caption{Changes in the diversity term when the diversity control parameter is changed.}\label{fig:diversity}
\end{figure}

In Figures~\ref{fig:sales} and \ref{fig:diversity}, the horizontal axes represent the value of $w$ in (\ref{form:qap}); the right hand side in the figures indicates that the larger the similarity between the similar items listed together, the larger the penalty.
In addition, the vertical axes in Figures~\ref{fig:sales} and \ref{fig:diversity} represent the values of $S({\bf x})$ and $D({\bf x})$, respectively. The average value of all solutions is approximately 0 because $s_{ij}$ and $f_{ii'}$ are normalized for each area.
Figures~\ref{sales_typical} and \ref{diversity_typical} depict plots of $S({\bf x})$ and $D({\bf x})$, respectively, for three typical areas X, Y, and Z from among the 10 areas for the problem with eight items.
Figures~\ref{sales_average} and \ref{diversity_average} are plotted by aggregating the typical values for the 10 areas, wherein the marker and the bar represent the average and standard error values, respectively.

Figure~\ref{sales_average} shows that the average of the sales term $S({\bf x})$ is higher than 0 for any $w$. In addition, $S({\bf x})$ gradually decreases as $w$ increases.
In other words, a trade-off relationship exists between increasing the sales term $S({\bf x})$ and ensuring that similar items are not placed adjacent to each other in the item list.
Furthermore, the decrease in the sales term value slows down as $w$ increases.

Figure~\ref{diversity_average} shows that when diversity is not considered for item listing, the diversity term $D({\bf x})$ is lower than the average value of 0 and increases with the increasing $w$.
Furthermore, the increase in the diversity term $D({\bf x})$ gradually slows down as $w$ increases.
Note that the behavior of $S({\bf x})$ and $D({\bf x})$ based on $w$ is the same whether or not the number of items is 6 or 8.

\begin{figure}[ht]
\centering
\includegraphics[width=8cm]{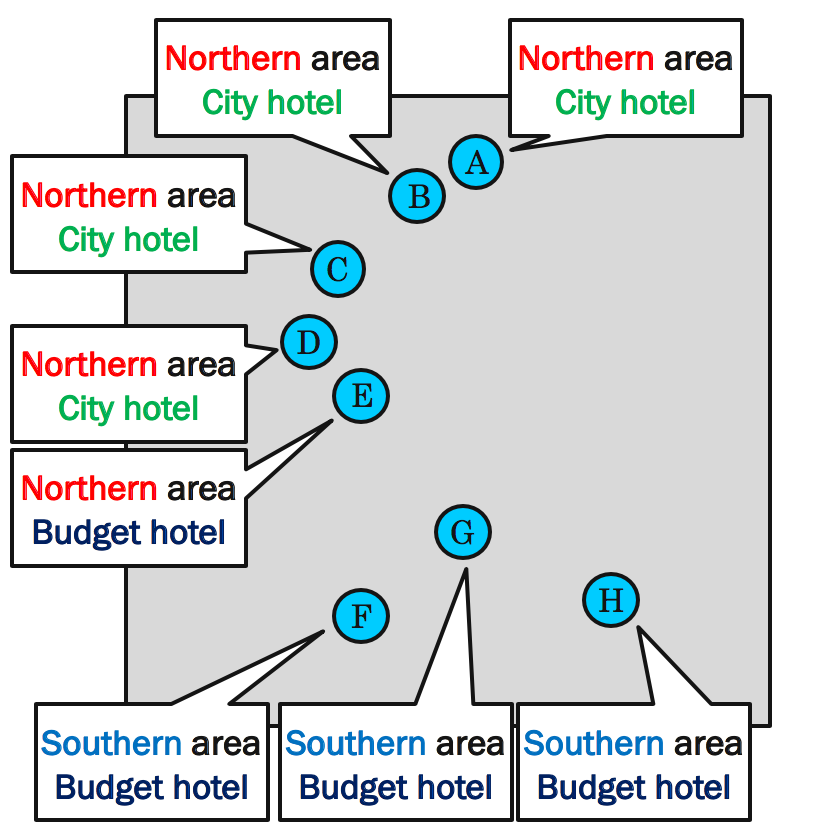}
\caption{Locations and type of a particular area’s hotels.}\label{fig:map}
\end{figure}

Figure \ref{fig:map} shows the positional relationships between the top eight hotels and their type in area X of Figures~\ref{sales_typical} and \ref{diversity_typical}.
The top eight hotels in area X particularly include four city hotels and four budget hotels, which can also be classified as in the north or south region of area X; hence, it was chosen as an example.

\begin{figure}[ht]
  \centering
  \includegraphics[width=16cm]{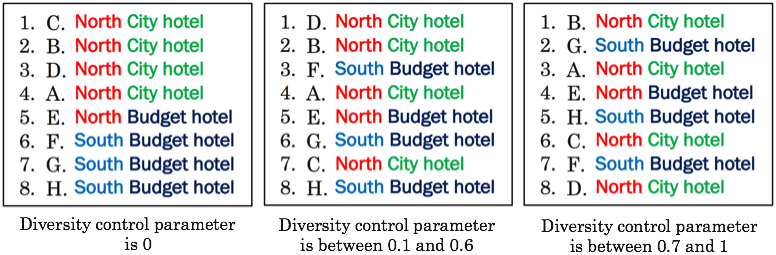}
  \caption{A particular area's item lists when the diversity control parameter $w$ is between 0 and 1.}\label{fig:140200}
\end{figure}

Figure~\ref{fig:140200} represents the actual hotel lists in area X obtained by solving (\ref{form:qap}) for different values of $w$.
The top five hotels in the northern area are consecutively listed when $w$ is $0$. The city hotels are listed in succession among the top four; thus, the list is biased.
In contrast, only the top two hotels are the same in terms of both area and hotel type when $w$ is between 0.1 and 0.6, that is, when diversity is considered.
Furthermore, no similarity exists between the consecutively listed hotels both in terms of area and hotel type when $w$ is larger than 0.7.
Thus, despite not using semantic information (e.g., area and hotel type) in the actual calculations, a semantically diverse item list was created using the number of co-browsers as the similarity measure.

\subsection{Performance Evaluation of Problem Decomposition}
We compared the method of extracting partial problems by \texttt{qbsolv} and our method of extracting partial problems considering the problem structure of the item list.
Our experiments were performed by comparing the objective values of (\ref{form:qap_qubo}) when solving the problem of 12, 16, 20 and 24 items using each method.

Table \ref{tab:decomp_objective_val} lists the average of the objective values for the 10 areas obtained by solving the problem for different numbers of items and the gap between the objective values of the proposed and original \texttt{qbsolv} methods. 
The objective of the problem was to minimize the objective value, indicating that the smaller the objective value, the better the performance.
The gap between the objective values of the proposed and original \texttt{qbsolv} methods increased as the number of items increased. 
The effectiveness of the proposed method also increased.
We conducted a Wilcoxon signed-rank test on the difference between the objective values of the original \texttt{qbsolv} and proposed methods (Bonferroni correction was performed on the number of items).
Consequently, the null hypothesis in this case was rejected at a significance level of 5\% (p-value = 0.020).

\begin{table}[ht]
\centering
\caption{Comparison of the objective values for problem decomposition.}\label{tab:decomp_objective_val}
\begin{tabular}{rrrr} \hline
   & \multicolumn{2}{c}{Methods} & \\
Number of items & \texttt{qbsolv} & Proposed method & Gap \\ \hline
12 & -160.038 & -160.337 & -0.299 \\ 
16 & -268.777 & -270.176 & -1.399 \\
20 & -391.428 & -393.051 & -1.623 \\
24 & -505.634 & -509.266 & -3.632 \\ \hline 
\end{tabular}
\end{table}

In terms of application, our proposed method is not only limited to the item list optimization problem, but can also be widely applied to other problems involving similar constraints, such as the assignment problem (\ref{form:ap}).
For example, our method can be applied to the traveling salesman problem, which typically includes two constants $ A, B $ along with the following penalty terms:
\begin{align}
A \sum_{i \in I}\left( \sum_{j \in J}x_{ij}-1\right) ^2  + B \sum_{j \in J}\left( \sum_{i \in I}x_{ij}-1\right) ^2. \nonumber
\end{align}

\section{Conclusion}
This study proposed a method of creating item lists on an e-commerce website as a QAP considering item sales and diversity.
We converted the QAP to a QUBO such that it can be directly solved by the quantum annealer, D-Wave 2000Q.
Direct manipulation to solve the resulting QUBO was not possible in the case with a large number of items because of the limited number of qubits available in the current version of the quantum annealer and the restriction on specifying connections between the qubits.
Therefore, we proposed a decomposition technique exploiting the structure of the problem.
The original large problem was divided into several subproblems, which can eventually be solved by D-Wave 2000Q individually.
Our experiments using actual real-world data demonstrated the efficiency of our proposed approach.
A remarkable observation made from the experimental results was that the output item list changed based on the diversity control parameter.
Our formulation led to the antiferromagnetic Ising model with a random field.
The resulting lists were ``aligned" along the random field when the diversity control parameter was small.
In contrast, increasing the diversity control parameter eliminated the order in the item list and introduced diversity.

However, note that our research has some limitations.
The item list created by our method will not work well when there are insufficient data to calculate the sales $s_{ij}$ and similarity $f_{ii'}$; because of which, no reliable estimate values existed. 
Having access logs when items were placed at various positions was important in obtaining good estimate values for $s_{ij}$ and $f_{ii'}$.
The determination of the diversity consideration parameter $w$ was also a limitation. 
$w$ depends on the scale of diversity a customer of an e-commerce service wants; thus, it should be adjusted by changing $w$ and monitoring performance, which is cumbersome to implementation in real-world scenarios.

As the number of qubits available in the quantum annealer increases in the future, our method for division of a large-sized problem into smaller subproblems will become more useful.
Our experiments clearly showed that our method performed better than \texttt{qbsolv} when the size of the problems became large.
The structure of our problem is similar to the traveling salesman problem and the scheduling problem in that it takes the form of a QUBO with two quadratic functions based on two constraints (i.e., one for distance or time and the other for list locations or tasks).

Thus, our method has a wide range of applications involving optimization problems that can be solved via QUBO formulation, such as those using the QA, D-Wave 2000Q, and other types of QUBO solvers.
Our results indicate that not only the evolution of hardware devices, but also the development of better software based on the structure of problems are essential for future QA applications.

\section*{Conflict of Interest Statement}
The authors declare that the research was conducted without any commercial or financial association that could be construed as potential conflicts of interest.

\section*{Author Contributions}
NN contributed to the conception and design of the study, performed all the experiments, and wrote the first draft of the manuscript. KT implemented the program of our problem decomposition method. MO and MJM  contributed to manuscript revision. All authors discussed this study, and then reviewed and approved the final version of the manuscript.

\section*{Acknowledgments}
The authors thank Recruit Lifestyle Co., Ltd. and Recruit Communications Co., Ltd. for its support in this exploratory research project.

\section*{Data Availability Statement}
The raw data supporting the conclusions of this manuscript will be made available by the authors, without undue reservation, to any qualified researcher.

\end{document}